\documentclass{epl}
\usepackage{amssymb}

\newcommand{\bq}{\begin{equation}}
\newcommand{\eq}{\end{equation}}
\newcommand{\bqn}{\begin{eqnarray}}
\newcommand{\eqn}{\end{eqnarray}}
\newcommand{\nb}{\nonumber}
\newcommand{\lb}{\label}

\title{On curvature coupling and quintessence fine-tuning}
\author{Yungui Gong\inst{1,2}\thanks{E-mail: \email{gongyg@cqupt.edu.cn}} \and Anzhong Wang\inst{2}
\and Yuan-Zhong Zhang\inst{3,4}}
\institute{\inst{1} College of Electronic Engineering, Chongqing
University of Posts and Telecommunications, Chongqing 400065,
China\\
\inst{2} CASPER, Physics Department, Baylor University,
Waco, TX 76798, USA\\
\inst{3} CCAST (World Laboratory), P.O. Box 8730, Beijing 100080, China\\
\inst{4} Institute of Theoretical Physics, Chinese Academy of Sciences,
P.O. Box 2735, Beijing 100080, China}
\pacs{98.80.Cq}{}
\pacs{98.80.Es}{}
\begin{document}
\maketitle

\begin{abstract}
We discuss the phenomenological model in which the potential energy of the
quintessence field depends linearly on the energy density of the
spatial curvature. We find that the pressure of the scalar field takes a different
form when the potential of the scalar field also depends on the scale factor and the energy
momentum tensor of the scalar field can be expressed as the form of a perfect fluid.
A general coupling is proposed to explain the
current accelerating expansion of the Universe and solve the
fine-tuning problem.
\end{abstract}

The astronomical observations tell us that the Universe is currently expanding in an acceleration
phase, while it was in a deceleration phase in the near past \cite{wmap,acc}. The accelerating
expansion of the Universe suggests
that 70\% of the total matter energy density comes from dark energy (DE) whose pressure is
negative. The simplest candidate of DE is the cosmological constant. Due to the
unusually small value of the cosmological constant given by observations, other DE
models are explored, such as the quintessence models \cite{quint} and the holographic DE models
\cite{hldark,intde}.
For a review of DE models, please see Ref. \cite{review} and references therein.

The quintessence models employ a scalar field that slowly rolls down its potential well to explain
the accelerating expansion. In the quintessence model, the potential energy
dominates the energy component, so the current value of the potential energy
must be of the order of the current critical energy density $3m_p^2 H^2_0=8.1h^2\times10^{-47}$
GeV$^4$, where the reduced Planck mass $m_p=(8\pi G)^{-1/2}=1.221\times 10^{18}$ GeV, the current
Hubble parameter $H_0=100h$ km/s/Mpc and $h=0.72\pm 0.08$ \cite{obsh}. It is not natural to
get the small potential energy for the quintessential scalar field, this is the
fine-tuning problem. Recently, Fran\c{c}a proposed a phenomenological model in which
the quintessential potential depends linearly on the curvature energy density $\rho_k=3m_p^2 k/a^2$
to solve the fine-tuning problem of DE model \cite{franca}. The existence of a small
curvature is consistent with the observation \cite{wmap}.
In this letter, we assume that the quintessential
potential depends linearly on $k/a^n$ and start from the equation of motion of the scalar field
to explain the accelerating expansion and solve the fine-tuning problem.

To begin with, we consider a general functional form of the potential
energy $U(\phi,a)$ of a homogeneous scalar field $\phi$.
By using the Robertson-Walker metric
$$ds^2=-dt^2+a^2(t)\left[\frac{dr^2}{1-kr^2}+r^2(d\theta^2+r^2d\varphi^2)\right],$$
we get the effective action for the
gravitational and quintessence fields as
\begin{equation}
\label{lag1}
\mathcal{S}_{eff}=\int dt a^3\left[3m_p^2\left(H^2-\frac{k}{a^2}\right)
-\frac{1}{2}\dot{\phi}^2+U(\phi,a)\right]+\mathcal{S}_o,
\end{equation}
where $\mathcal{S}_o$ denotes the action for the matter and radiation fields.
Because of the dependence of the potential on the scale factor $a$,
the Friedmann equation will change.
From the action (\ref{lag1}), varying, respectively,  $a$ and $\phi$,
we get the following equations of motion
\begin{equation}
\label{cos2}
2\frac{\ddot{a}}{a}+H^2+\frac{k}{a^2}=-\frac{1}{m_p^2}\left[p_m+p_r+
\frac{1}{2}\dot{\phi}^2-U(\phi,a)-\frac{a}{3}\frac{\partial U(\phi,a)}{\partial a}\right],
\end{equation}
\begin{equation}
\label{quint1}
\ddot{\phi}+3H\dot{\phi}+ \frac{\partial U(\phi,a)}{\partial\phi} = 0,
\end{equation}
where $p_m$ is the pressure for  the matter, $p_r$ the pressure for radiation,
and the Hubble parameter is defined as $H \equiv \dot{a}/a$.

It should be noted that due to the explicit dependence of the potential $U(\phi, a)$
on the metric coefficients, the energy-momentum tensor $T^{\mu\nu}_{\phi}$ no longer
takes its usual form,
\begin{equation}
\label{eqa}
T_{\mu\nu}^{\phi} \equiv 2(-g)^{-1/2} \frac{\delta (\sqrt{-g}{\cal{L}}_{\phi})}{\delta g^{\mu\nu}}
\not= \phi_{,\mu} \phi_{,\nu} - g_{\mu\nu}{\cal{L}}_{\phi},
\end{equation}
where ${\cal{L}}_{\phi} \equiv \frac{1}{2}(\nabla \phi)^{2} + U(\phi, a)$. Consequently, now
we cannot write it in the form,
\bq
\lb{eqb}
T_{\mu\nu}^{\phi} = (\rho_{\phi} + p_{\phi})u_{\mu}u_{\nu} + p_{\phi} g_{\mu\nu},
\eq
with $u_{\mu} = {\left|\phi_{,\alpha}\phi^{,\alpha}\right|^{-1/2}}{\phi_{,\mu}}$
and
\bq
\lb{eqc}
\rho_{\phi} = \frac{1}{2} \left(\nabla\phi\right)^{2} + U(\phi, a),\;\;
p_{\phi} = \frac{1}{2} \left(\nabla\phi\right)^{2} - U(\phi, a).
\eq
This point can be seen clearly from our discussions to be given below.
Since there is no such kind of coupling with the matter and radiation fields, as one
can see from the action (\ref{lag1}), we still have
\bq
\lb{eqd}
T^{0}_{\mu\nu} = T^{m}_{\mu\nu} + T^{r}_{\mu\nu} =
\left(\rho_{m} + \rho_{r} + p_{m} + p_{r}\right) u_{\mu}u_{\nu}
+ \left(p_{m} + p_{r}\right) g_{\mu\nu},
\eq
where $p_{m} = 0$ for the dust matter and $p_{r} = \rho_{r}/3$ for the radiation. Without
loss of generality, in the following we shall consider only these two different kinds of
matter fields. Then, the conservation laws $T^{m}_{\mu\nu; \lambda} g^{\nu\lambda} =
0 = T^{r}_{\mu\nu; \lambda} g^{\nu\lambda}$ yield
\bq
\lb{eqe}
{d}\left(\rho_{m} a^{3}\right) = 0, \quad
{d} \left(\rho_{r} a^{3}\right)
+ p_{r} {d}\left(a^{3}\right) = 0.
\eq
If we multiply equation (\ref{cos2}) by $H$ and use the above equations and equation (\ref{quint1}), we get
\bqn
H\left(2\frac{\ddot{a}}{a}+H^2+\frac{k}{a^2}\right)&=&3H\left(H^2+\frac{k}{a^2}\right)
+\frac{d}{dt}\left(H^2+\frac{k}{a^2}\right)\nonumber\\
&=&\frac{1}{3m_p^2}\left[\dot{\rho_m}+3H\rho_m+\dot{\rho_r}+3H\rho_r+\dot{\phi}\ddot{\phi}
+\frac{\partial U}{\partial \phi}\dot{\phi}+\frac{\partial U}{\partial a}\dot{a}+
\frac{3}{2}H\dot{\phi}^2+3HU\right]\nonumber\\
&=&\frac{3H}{3m_p^2}\left[\rho_m+\rho_r+\frac{1}{2}\dot{\phi}^2+
U\right]+\frac{1}{3m_p^2}\frac{d}{dt}\left[\rho_m+\rho_r+\frac{1}{2}\dot{\phi}^2+
U\right].
\eqn
So
\begin{equation}
\label{cos3}
H^2+\frac{k}{a^2}=\frac{1}{3m_p^2}\left[\rho_m+\rho_r+\frac{1}{2}\dot{\phi}^2+
U(\phi,a)\right],
\end{equation}
\begin{equation}
\label{cosacc2}
\frac{\ddot{a}}{a}=-\frac{1}{6m_p^2}\left[\rho_m+2\rho_r+2\dot{\phi}^2-2U(\phi,a)
-a\frac{\partial U(\phi,a)}{\partial a}\right],
\end{equation}
where the matter energy density $\rho_m=\rho_{m0}(a_0/a)^3$,
the radiation energy density $\rho_r=\rho_{r0}(a_0/a)^4$, and the subscript $0$ means that
the variable is evaluated at the present time.
Comparing equations (\ref{cos2}), (\ref{cos3}) and (\ref{cosacc2}) with those in standard cosmology, we
find that
\bqn
\lb{eqf}
\rho_\phi &=& \frac{1}{2}\dot{\phi}^2 + U(\phi,a),\nb\\
 p_\phi &=& \frac{1}{2} \dot{\phi}^2 - U(\phi,a)
- \frac{1}{3}a \frac{\partial U(\phi, a)}{\partial a},
\eqn
which are clearly different from those given by equation (\ref{eqc}). As mentioned above, the reason
is exactly because of  the dependence of the potential on the scale factor $a(t)$.
With these new definitions, equation (\ref{cosacc2}) can be written in its familiar form,
\begin{equation}
\label{cosacc3}
\frac{\ddot{a}}{a}=-\frac{1}{6m_p^2}(\rho_m+2\rho_r+\rho_\phi+3p_\phi),
\end{equation}
while the Klein-Gordon equation (\ref{quint1}) reads
\begin{equation}
\label{encons}
\dot{\rho}_\phi+3H(\rho_\phi+p_\phi)=0.
\end{equation}

If we take \bq \lb{pot} U(\phi,a)=\frac{ k V_0}{a^n} V(\phi), \eq
with $V_0=3m^2_p$ and $V_0 k/(3m_p^2 H_0^2)=\Omega_{k0}=0.02$ \cite{wmap}, and $n$
being a positive number, we get
$p_\phi=\dot{\phi}^2/2-(1-n/3)U(\phi,a)$. So
$\rho_\phi+3p_\phi=2\dot{\phi}^2+(n-2)U(\phi,a)$. It is obvious
that $\rho_\phi+3p_\phi\ge 0$ if $n\ge 2$. Therefore from equation
(\ref{cosacc3}), we see that there is no accelerating expansion of
the Universe if $n\ge 2$. Note that the potential with $n = 2$ is
exactly the one used in \cite{franca}, in which it was shown that
the cosmic coincidence problem can be solved with such a coupling,
while the universe is still accelerating \cite{franca2}. Clearly,
this result is different from what we obtained here.  The main
reasons, as explained by Fran\c{c}a in \cite{franca2}, are the
following: Instead of starting with the effective action
(\ref{lag1}), Fran\c{c}a started with the energy-momentum tensor
$T^{\phi}_{\mu\nu}$ given by \bq \lb{eqg} T_{\mu\nu}^{(\phi, Fr.)}
= \phi_{,\mu} \phi_{,\nu} -
\frac{1}{2}g_{\mu\nu}\left(\phi_{,\alpha}\phi^{,\alpha} -
2U(\phi,a)\right), \eq from which it can be shown that the
conservation law $T^{(\phi, Fr.)}_{\mu\nu; \lambda} g^{\nu\lambda}
 = 0$ gives
\bq \lb{eqh} \ddot{\phi} + 3H\dot{\phi} + \frac{\partial U(\phi,
a)}{\partial \phi} + \frac{da}{d\phi}  \frac{\partial U(\phi,
a)}{\partial a} = 0, \eq
which is different from
equation (\ref{quint1}). Assuming that (a) the standard Friedmann
equation holds, and (b) $\rho_{\phi}$ and $p_{\phi}$ are related
to the scalar field by equation (\ref{eqc}), Fran\c{c}a was able to
show that the acceleration can also be written in the exact form
of equation (\ref{cosacc3}). Then, from equation (\ref{eqc}) one can show that
$\rho_{\phi} + 3p_{\phi} = 2(\dot{\phi}^{2} - U)$,  from which we
can see that it becomes possible that the universe is accelerating
whenever $\dot{\phi}^{2} < U$.

From the above analysis we see that in this letter we are actually considering a completely
different case from that of Fran\c{c}a \cite{franca,franca2}, although the potentials used in these
two cases are similar. In particular, to get the energy-momentum tensor given by equation (\ref{eqg}),
the effective action will be quite different from that of equation (\ref{lag1}).

Once the above points are made clear, following \cite{franca} we
consider  the potential $V(\phi)=\exp(\lambda\phi/m_p)$. It is interesting to note that
the potential $U(\phi,a)$ with curvature coupling also gives the tracking solution in which the
scalar field $\phi$ tracks the background matter field and $\Omega_\phi=\rho_\phi/(3m_p^2 H^2)=
(3+3w_b-n)(6-n)/6\lambda^2$, here $w_b$ is the equation of state parameter for the background
matter field.

Now we change
the variable from $t$ to $u=\ln(a/a_0)$, denote the derivative
to the new variable $u$ by $\prime$ and take $y=\phi/m_p$, equations (\ref{cos3}) and
(\ref{quint1}) become
\begin{equation}
\label{curvquint1}
\left(\frac{H}{H_0}\right)^2=\frac{\Omega_{m0}e^{-3u}+\Omega_{r0}e^{-4u}
-\Omega_{k0}e^{-2u}+\Omega_{k0}a^{2-n}_0e^{-n u+\lambda y}}{1-y'^2/6},
\end{equation}
and
\begin{eqnarray}
\label{curvquint2}
\left(\frac{H}{H_0}\right)^2 y''+\left(\frac{3}{2}\Omega_{m0}e^{-3u}+\Omega_{r0}e^{-4u}
-2\Omega_{k0}e^{-2u}+\frac{6-n}{2}\Omega_{k0}a^{2-n}_0e^{-n
u+\lambda y}\right)y'+\nonumber\\
 3\lambda \Omega_{k0}a^{2-n}_0 e^{-n
u+\lambda y}=0,
\end{eqnarray}
where
$\Omega_{m(r,k)}=\rho_{m(r,k)}/(3m_p^2 H^2)$.
At late times, the
dark energy component will dominate the Universe, so we can
neglect the contributions of matter, radiation and curvature to
the energy density. For a scaling solution $\phi''=0$, equation
(\ref{curvquint2}) gives us the solution
\begin{equation}
\label{scalsol}
y'=\frac{-6\lambda}{6-n}.
\end{equation}
Therefore, we get
\begin{equation}
\label{wde}
w_\phi=\frac{p_\phi}{\rho_\phi}=-1+\frac{n}{3}+\frac{2\lambda^2}{6-n}.
\end{equation}
For $\lambda^2<(3-n)(6-n)/6$, the above scaling solution is also an attractor solution. At present,
the attractor is not reached yet. Therefore we need to fine-tune the initial conditions of the
scalar field itself so that we get $\Omega_{m0}=0.28\pm 0.04$ and $w_{\phi 0}\le -0.8$ \cite{gong05}. However,
the fine-tuning of the initial conditions are different from the fine-tuning of the energy
scale $V_0$ of the potential. Note that we take $V_0=3m_p^2$ in this letter, so we solve the
fine-tuning problem of the energy scale.

From the definitions, we get
\bq
\lb{omphi}
\Omega_\phi=\frac{1}{6}y'^2+\Omega_{k0}a_0^{2-n}\left(\frac{H_0}{H}\right)^2e^{-nu+\lambda y},
\eq
and
\bq
\lb{wphi}
w_\phi=-1+\frac{n}{3}+\frac{(2-n/3)y'^2}{6\Omega_\phi}>-1+\frac{n}{3}.
\eq
So we require $n\le 0.6$ in order to satisfy the observational constraint $w_{\phi 0}\le -0.8$.
For the case $n=0.6$, the possible value for the coupling constant $\lambda$ is $\lambda=0$.
For $\lambda=0$, the scalar field effectively becomes the DE model with $w_\phi=-1+n/3$.

\begin{figure}[hbtp]
\begin{center}
\includegraphics[width=10cm]{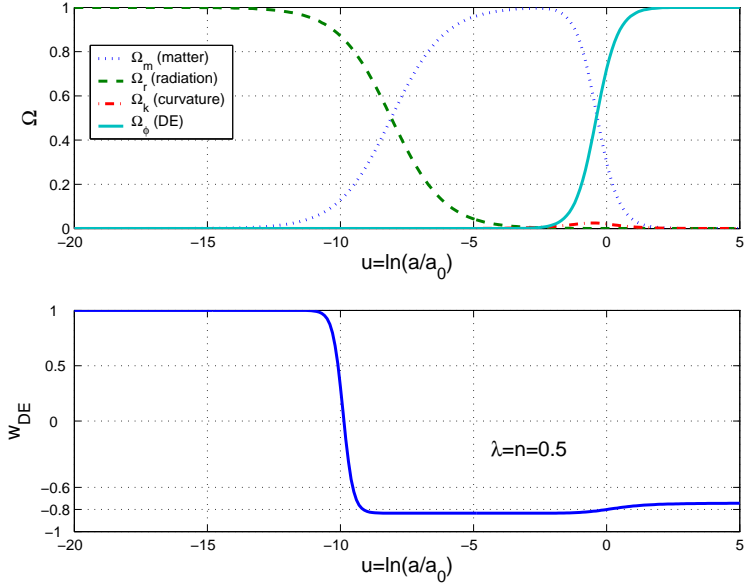}
\end{center}
\caption{The evolutions of the energy densities $\Omega_m$, $\Omega_r$, $\Omega_k$ and $\Omega_\phi$
and the equation of state parameter $w_\phi$ of the quintessence field.}
\label{fig1}
\end{figure}

\begin{figure}[hbtp]
\begin{center}
\includegraphics[width=10cm]{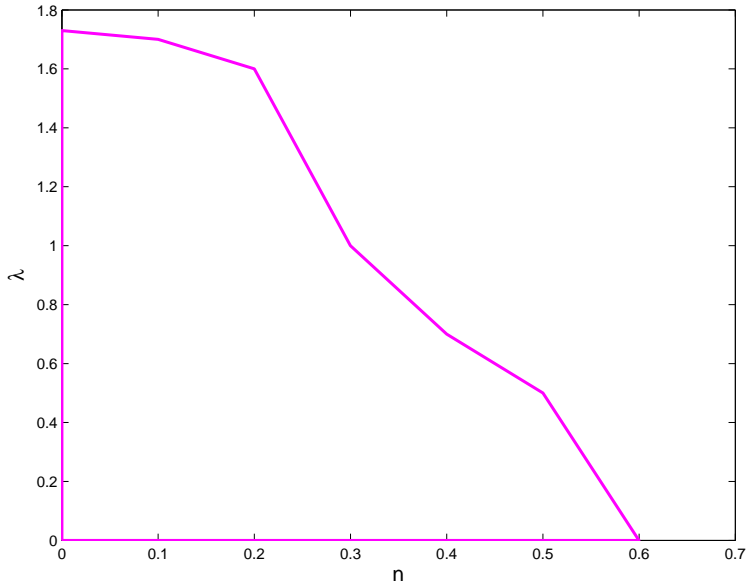}
\end{center}
\caption{The allowed parameter space from observational constraints.}
\label{fig2}
\end{figure}

To get the evolutions of the DE energy density and the Universe,
we solve equations (\ref{curvquint1}) and (\ref{curvquint2})
numerically starting from the epoch of nucleosynthesis when $a_i/a_0\sim 10^{-10}$. Here
the subscript $i$ denotes the epoch of nucleosynthesis. Since we solve
the equations starting from the epoch $a_i$, we rewrite equations (\ref{curvquint1}) and (\ref{curvquint2})
as
\begin{equation}
\label{curvquint3}
\left(\frac{H}{H_i}\right)^2=\frac{\Omega_{mi}e^{-3v}+\Omega_{ri}e^{-4v}
-\Omega_{ki}e^{-2v}+\Omega_{ki}a^{2-n}_ie^{-n v+\lambda y}}{1-y'^2/6},
\end{equation}
and
\begin{eqnarray}
\label{curvquint4}
\left(\frac{H}{H_i}\right)^2 y''+\left(\frac{3}{2}\Omega_{mi}e^{-3v}+\Omega_{ri}e^{-4v}
-2\Omega_{ki}e^{-2v}+\frac{6-n}{2}\Omega_{ki}a^{2-n}_ie^{-n
v+\lambda y}\right)y'+\nonumber\\
 3\lambda \Omega_{ki}a^{2-n}_i e^{-n
v+\lambda y}=0,
\end{eqnarray}
where $v=\ln(a/a_i)=\ln(a_0/a_i)+u$, $\Omega_{mi}/\Omega_{ri}=(a_i/a_0)(\Omega_{m0}/\Omega_{r0})$,
$\Omega_{ki}/\Omega_{ri}=(a_i/a_0)^2(\Omega_{k0}/\Omega_{r0})$. When the
initial conditions are specified, equations (\ref{curvquint1}) and (\ref{curvquint2})
can be solved numerically.
For example, if $\lambda=n=0.5$, we get $w_{\phi 0}=-0.801$ and
$\Omega_{m0}=0.298$. If $\lambda=0.5$ and $n=0.1$, we get $w_{\phi 0}=-0.937$ and $\Omega_{m0}=0.297$.
Since $\Omega_\phi$ is very small at early time, from equation (\ref{omphi}) we know
that $y'$ is small and so the scalar field changes very slowly. The observational constraint
$\Omega_{m0}\sim 0.28$ gives the initial condition
$\Omega_{k0}(a_0/a_i)^{2-n}\exp(\lambda y_i)\sim \Omega_{\phi 0}$.
The other initial condition is taken to be $y'_i=-0.8$.
The evolutions of the energy densities
$\Omega_m$, $\Omega_r$, $\Omega_k$ and $\Omega_\phi$ are shown in the top panel of figure \ref{fig1}. The evolution of
the equation of state parameter $w_\phi=p_\phi/\rho_\phi$ of the quintessence field is shown
in the bottom panel of figure \ref{fig1}.
From figure \ref{fig1}, we see that DE dominates the Universe now and it has the correct
equation of state to give the acceleration.
By using the observational constraint, we also obtain the constraints on the parameters $\lambda$ and $n$
in figure \ref{fig2}. From figure \ref{fig2}, we see that the bigger $n$ is, the smaller $\lambda$ is allowed.
This can be understood from equation (\ref{curvquint2}). If $\lambda$ is bigger, then $y''$ becomes more negative
when the scalar field dominated the Universe. So $y'$ decreases faster and takes more negative value. Therefore from
equation (\ref{wphi}), we see that $n$ must be smaller to satisfy the constraint of $w_{\phi}$.

In summary, we analyze the curvature dependence model
proposed by Fran\c{c}a which solves the fine-tuning
problem of DE model. We find that the definitions of the energy density and pressure
of the scalar field should be changed in order that the Klein-Gordon equation of the scalar field
is equivalent to the energy conservation equation for the scalar field if the scalar field potential
also depends on the scale factor. We then modify
the curvature dependence of the quintessence potential from $k/a^2$ to $k/a^n$. In
the phenomenological model in which the quintessence potential depends linearly on $k/a^n$
for $n\le 0.6$, we find
that the fine-tuning problem of the quintessence potential can be solved too.
We would like to stress that the model is also a pure phenomenological one. The exact
form of the non-minimal coupling of the scalar field to gravity is still unknown.

\acknowledgments

Y. Gong thanks U. Fran\c{c}a for many fruitful discussions.
Y. Gong is supported by Baylor University, NNSFC under grants 10447008 and 10575140, and SRF for ROCS,
State Education Ministry of China. Y.Z.
Zhang's work was in part supported by NNSFC under Grant No.
90403032 and also by National Basic Research Program of China
under Grant No. 2003CB716300.


\end{document}